\documentclass[aps,prl,reprint]{revtex4-2}
\usepackage{graphicx,graphics}
\usepackage{epstopdf} 
\usepackage{amsmath,amssymb,amsfonts}
\usepackage{latexsym,verbatim}
\usepackage{bm}
\usepackage{bbold}
\usepackage{xcolor}
\usepackage[normalem]{ulem}
\usepackage[breaklinks=false,colorlinks,citecolor=blue,linkcolor=blue,urlcolor=blue]{hyperref}
\usepackage[abs]{overpic}
\usepackage{textcomp} 
\usepackage{mathtools}
\usepackage{braket}
\usepackage{soul}    

\newif\ifdraft
\drafttrue 

\definecolor{Mygrey}{gray}{0.80}
\definecolor{lteal}{rgb}{0.10,0.60,0.70}
\definecolor{dkred}{rgb}{0.80,0.10,0.00}
\definecolor{dgreen}{rgb}{0.0, 0.5, 0.0}

 \ifdefined\LightComments

\else

 \fi

\begin{document}

\title{Genuine multipartite entanglement from many-electron systems}

\author{Filippo Troiani$^1$}
\email{filippo.troiani@nano.cnr.it}
\author{Celestino Angeli$^2$}
\author{Andrea Secchi$^1$}
\author{Stefano Pittalis$^1$}
\affiliation{$^1$Centro S3, CNR-Istituto di Nanoscienze, I-41125 Modena, Italy}
\affiliation{$^2$Dipartimento di Chimica, Università di Ferrara, I-44100 Ferrara, Italy}

\begin{abstract}
{\color{black}
We demonstrate that, contrary to common wisdom, genuine multipartite entanglement (GME) can be abundantly generated  from simple non-correlated many-electron states. We show that the extracted GME can be maximized via spin-independent transformations derived from the quantum Fourier transform. We further demonstrate the possibility of maximizing the GME through localized orbitals in a variety of  realistic systems and correlated states. Towards the exploitation of potentially useful entanglement, we rationalize system-specific and universal features of the extracted GME.}
\end{abstract}

\date{\today}

\maketitle
{\em Introduction.}~Quantum entanglement is an essential resource in quantum technologies. Largely driven by such a motivation, the investigation of entanglement has been mainly focused on the case of qubits \cite{Amico00a,Horodecki09a}.
{\color{black} In the case of bosons and fermions --- which are neither distinguishable, nor individually addressable --- the definition, the quantification, and the exploitation of entanglement are currently under intense debate~\cite{Benatti20a,Sun22a}. 

Remarkably, indistinguishability of bosons was shown to represent a potential resource in a variety of applications, ranging from quantum communication to sensing, and can be exploited in the generation of diverse multipartite entangled states \cite{Mahdavipour24a,Piccolini_2025}}. 
The case of fermions is most relevant for understanding and harnessing entanglement in diverse systems, from atoms and molecules to materials~\cite{Schliemann01a, Li01a,Zanardi02a,Amico08a,Franca08a,Gigena15a,Franca11a}.
{\color{black} Here, a major distinction can be made between approaches that focus on correlations between particles or modes \cite{Benatti20a}.} 
Orbital entanglement \cite{Zanardi02a,Legeza2003,Rissler2006,Boguslawski2015} {\color{black} --- belonging to the latter class ---} has also found useful application in the domain of quantum chemistry \cite{Boguslawski2013,Szalay17a,Tenti24,Duperrouzel2015,Stein2016}, but cannot be regarded as a technological resource~\cite{Wiseman03a,Ding2021}.

\begin{figure}
\centering
\includegraphics[width=0.45\textwidth]{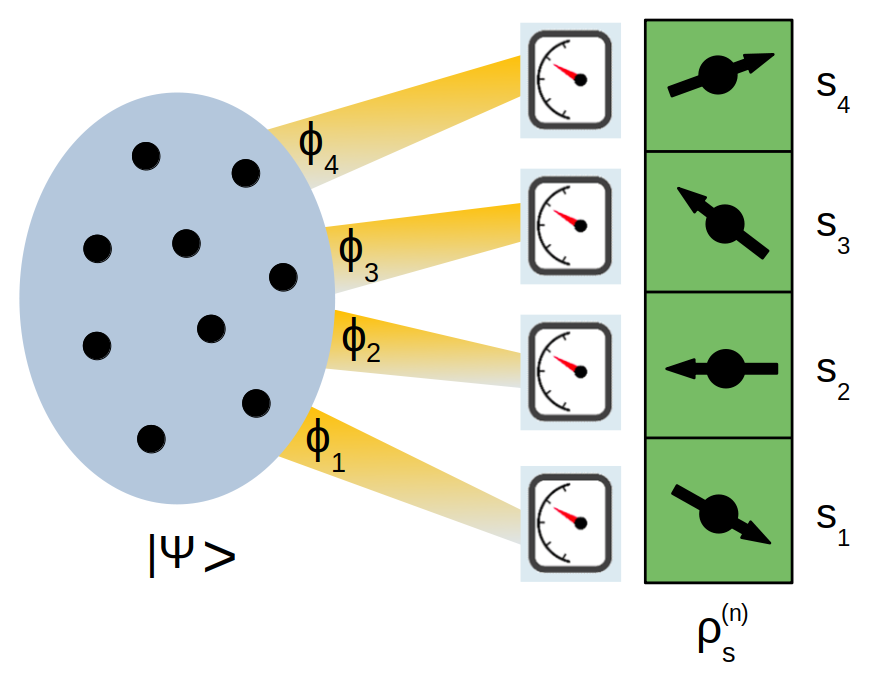}
\caption{Schematic representation of the considered procedure: from an $N$-electron system in a state $|\Psi\rangle$, one extracts an $n$-spin state $\rho_s^{(n)}$ ($n \le N$ in general, $N=9$ and $n=4$ in the figure). Each of the $n$ spins $s_i$ is labelled by the respective orbital $\phi_i$, whose physical meaning is discussed in the main text.
}
\label{fig1}
\end{figure}

{\color{black} In an alternative approach, one can reduce a fermionic to a spin (or qubit) system by freezing the spatial degrees of freedom, and using them for an unambiguous particles labeling \cite{Benatti20a}. Such an effective ``distinguishability" can be related to the generation of entanglement through particle detection.~\cite{Killoran14a,Bouvrie17a,LoFranco18a} 
This post-selects specific components of the original fermionic state, namely the ones that allow the multiple counting events, taking place along well-defined spatial modes (or orbitals). As a result,} 
the entanglement properties of the extracted $n$-spin state result both from the properties of the original $N$-electron state $|\Psi\rangle$, and from the set of orbitals $\phi_i$ that define the particle extraction 
(Fig. \ref{fig1}).

{\color{black} The detection of particles can be formally associated to Green's functions \cite{pilar2001elementary}. 
The formalism used for the derivation of the spin states can be introduced by referring to the simple $n=2$ case. The spin state of the two electrons labeled by (detected at)} the orbitals $\phi_i$ and $\phi_j$ is described by the $2$-Body Reduced Density Matrices ($2$BRDM)~\cite{Vedral03a,Shi04a,Oh04a}, whose expression reads:
\begin{gather}\label{eqn:2BRDM}
    \Gamma_{ij;\sigma_i,\sigma_j,\sigma_i',\sigma_j'}^{(2)} \!=\! \langle \Psi | \hat{d}_{i,\sigma_i'}^\dagger \hat{d}_{j,\sigma_j'}^\dagger  \hat{d}_{j,\sigma_j} \hat{d}_{i,\sigma_i} |\Psi\rangle\;.
\end{gather}
Here, $\hat{d}_{i,\sigma_i}$ ($\hat{d}_{i,\sigma_i}^{\dagger}$) are fermionic operators that annihilate (create) an electron in the orbital $\phi_i$ with spin 
$\sigma_i  = \uparrow, \downarrow$.
The 2BRDMs can thus be regarded as the matrix representations of spin states, {\color{black} with row and column indices given by $(\sigma_i',\sigma_j')$ and $(\sigma_i,\sigma_j)$, respectively. Analogously for the $n$BRDMs discussed below}. 

This approach has been applied to the study of spin entanglement in a Fermi gas~\cite{Vedral03a,Oh04a} --- a minimalist model for electrons in metals.
While the single-particle orbitals of the Fermi gas are fully delocalized in space (plane waves), the extracted spins can be labeled through the eigenfunctions of the position operator. The resulting spin entanglement of the extracted particles monotonically decreases with the interparticle distance, and is nonzero as long as such distance is sufficiently small compared to the Fermi wave length~\cite{Vedral03a,Oh04a}.
This analysis has been generalized to atomic and molecular systems, where the length scale characterizing the decay of spin entanglement is inhomogeneous --- clearly reflecting the presence of atomic shells and molecular bonds --- and is directly related to key ingredients in electronic structure theory~\cite{Pittalis15a}.

In this Letter, we turn to investigate the {\em genuine} multipartite entanglement (GME), which involves more than two parties. While an early work has claimed that no GME can be extracted from a Fermi gas~\cite{Lunkes05a} {\color{black} and, more generally, from uncorrelated closed-shell states,} we demonstrate here that such an extraction is instead possible. {\color{black} Close to maximal GME is also shown to result from  correlated eigenstates of prominent interacting electron models, ranging from Hubbard to atomistic Hamiltonians. In all these cases, } the generation of GME is mediated by single-particle orbitals {\color{black} that possess a clear physical meaning and are spatially localized: a feature that} suggests the possibility of extracting the particles from well-defined locations. 

The $n$-spin state extracted from the orbitals $\phi_1,\dots,\phi_n$ is formally given by the $n$BRDM:
\begin{gather}\label{eqn:nBRDM}
    \Gamma^{(n)}_{{\bf\sigma},{\bf\sigma}'} \!=\! \langle \Psi | \,\hat{d}_{n,\sigma_n'}^\dagger \dots \hat{d}_{1,\sigma_1'}^\dagger \hat{d}_{1,\sigma_1} \dots \hat{d}_{n,\sigma_n}\, |\Psi\rangle\;.
\end{gather}
Here, the indices in the subscript of the $n$RDM that specify the orbitals run from $1$ to $n$ and have been omitted for simplicity, while $\sigma$ and $\sigma'$ are $n$-dimensional vectors, whose components $\sigma_i=\uparrow,\downarrow$ specify the spin state. If the number of extracted electrons coincides with the overall number of particles ($n=N$), the extracted spin state coincides with a pure state (see Sec. III of the Supplemental Material \cite{SM}).

When the $n$BRDM corresponds, up to a normalization factor, to the projector on a pure spin state $|\Phi_s\rangle$, the GME can be quantified via the GME concurrence~\cite{Ma11b,Huber2010}:
\begin{align}\label{eq:GMEC}
    \mathcal{C}_{\rm GME} (|\Phi_s\rangle) 
    \!=\! 2\min_{\xi\in\Xi} \left[ 1\!-\!{\rm Tr}\left( \rho^2_{A_{\xi}} \right) \right]^{\frac{1}{2}}
    \!\equiv\! \sqrt{ 2 S_L(\rho_{A_m}) }.
\end{align}
Here $\Xi$ represents the set of all possible bipartitions $\xi$ of the $n$-spin system into the subsystems $A_\xi$ and $B_\xi$, $\rho_{A_{\xi}}$ is the reduced density matrix of $A_{\xi}$, and $A_m$ is the subsystem with the lowest linear entropy $S_L$.

\begin{figure}[h!]
\centering
\includegraphics[width=0.45\textwidth]{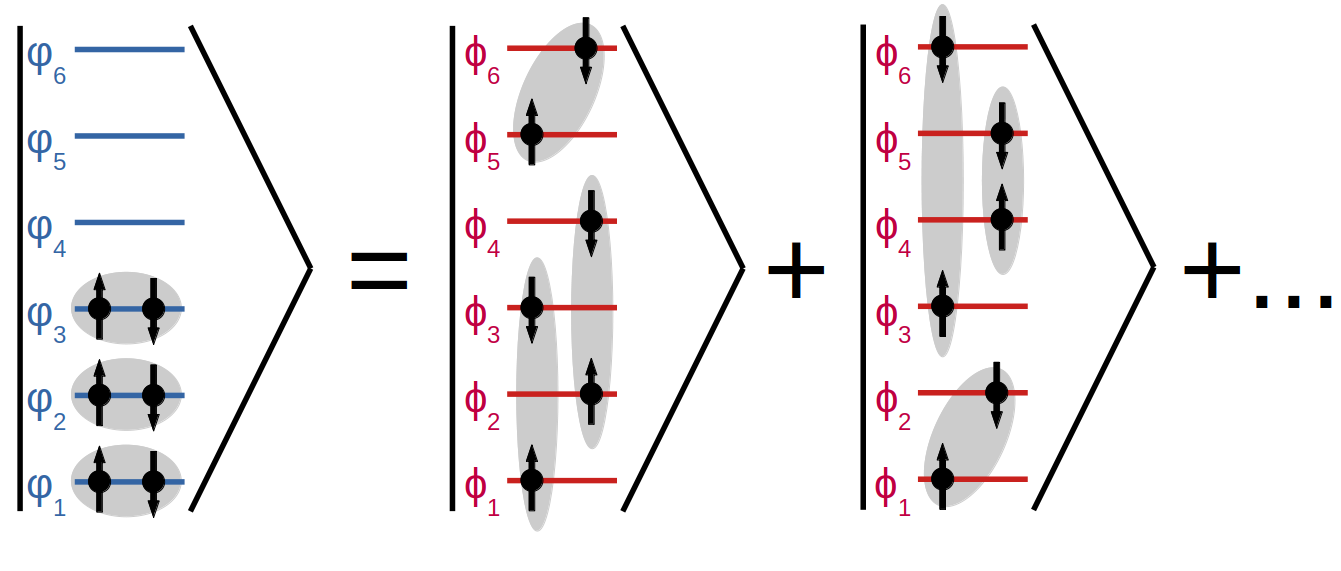}
\includegraphics[width=0.45\textwidth]{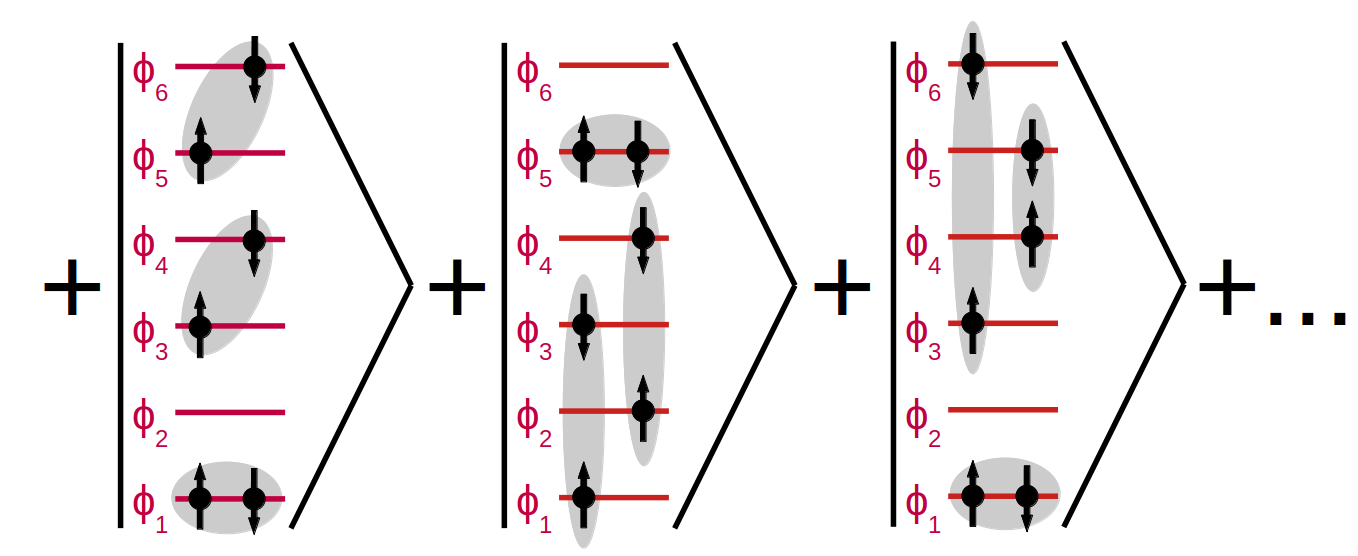}
\caption{{\color{black} In the basis $\{\varphi_i\}$ (blue levels), the state $|\Psi\rangle$ is defined as a single configuration, corresponding to the double occupation of the first $N/2$ orbitals. In the basis $\{\phi_j\}$ (red levels), $|\Psi\rangle$ is given by a linear superposition, with different coefficients, of many configurations. Each of these corresponds to the product of $N/2$ singlets (gray areas), either delocalized on a pair of orbitals, or localized on a single orbital.}}
\label{fig2}
\end{figure}

{\color{black} \em GME from single-Slater-determinants}.~To upgrade the case of the Fermi gas {\color{black} (see Sec. IV of the Supplemental Material \cite{SM}) and generalize it to realistic weakly correlated systems,} we consider the generic single-reference closed-shell state:
\begin{align}\label{eqn:Psi}
    \big| \Psi \big> = \left( \prod_{i = 1}^{N/2} \hat{c}^\dagger_{i,\uparrow} \hat{c}^\dagger_{i,\downarrow} \right) \big| {\rm vac} \big> \,,
\end{align}
where $\hat{c}^\dagger_{i,\sigma}$ creates an electron in the orbital $\varphi_i$ with spin $\sigma=\uparrow,\downarrow$. 
The $N$-electron state $|\Psi\rangle$ corresponds to a product of singlet two-electron states, each one {\em localized} on an orbital $\varphi_i$. Here, the term ``localization'' refers to the distribution of the spin among the orbitals --- not to their spatial extension (we consider this latter aspect below).

The relation between the orbitals $\varphi_i$ and those that define the particle extraction $\phi_i$ is given by a general spin-independent transformation. {\color{black} This is identified by an $N \times N$ unitary matrix $U$, whose elements define the transformation between the relevant} creation operators:
$
\hat{c}_{i,\sigma}^\dagger = \sum_{j=1}^{N} u_{ij}\, \hat{d}_{j,\sigma}^\dagger
$, where 
$\hat{d}^\dagger_{j,\sigma}$ 
creates an electron in the orbital $\phi_j$ with spin $\sigma$.
It is worth noticing that
\begin{gather}\label{eq:locsin}
\hat{c}^\dagger_{i,\uparrow}  \hat{c}^\dagger_{i,\downarrow} =   \sum_{j,k = 1}^{N} u_{ij} u_{ik}\, \hat{S}^{\dagger}_{j,k}\,,
\end{gather}
where we have introduced the {\it singlet creation operators}: 
\begin{align}\label{eqn:SCO}
    \hat{S}^{\dagger}_{j,k} \equiv \frac{1}{2} \left( \hat{d}^{\dagger}_{j, \uparrow} \hat{d}^{\dagger}_{k, \downarrow} - \hat{d}^{\dagger}_{j, \downarrow} \hat{d}^{\dagger}_{k, \uparrow}  \right)    \,.
\end{align}
These operators create a ``delocalized'' (over the two orbitals $\phi_j$ and $\phi_k$) spin singlet if $j \neq k$, and a localized one for $j = k$. In the orbital basis $\{\phi_i\}$, the state $|\Psi\rangle$ given in Eq.~\eqref{eqn:Psi} is thus a linear superposition of different configurations, each one given by the product of $N/2$ localized or delocalized singlets (Fig. \ref{fig2}).
{\color{black} Formally, this character of the $N$-electron state $|\Psi\rangle$ emerges from the combination of Eqs. (\ref{eq:locsin}) and (\ref{eqn:SCO}), which results in the expression:
\begin{align}
    \big| \Psi \big> = \left( \prod_{i=1}^{N/2} \sum_{j_i=1}^{N} \sum_{k_i=1}^{N} u_{ij_i} u_{ik_i}  \hat{S}^{\dagger}_{j_i, k_i} \right) \big| {\rm vac} \big>  \,.
    \label{Psi in terms of d}
\end{align}

We start by considering the case where the relation between the two sets of orbitals coincides with the quantum Fourier transform (QFT) \cite{Nielsen_Chuang_2010}:
$
u_{jk} = e^{-i\,2\pi\,jk/N} / \sqrt{N}
$.
The spin states defined by these $n$BRDMs are characterized by values of the GME that are close to the theoretical maximum, and monotonically increase with $n$ (Table \ref{tab:my_label}). Spin-pair entanglement, which is only present between nearest neighbors and decreases with $n$, is quantified by the reported values of the concurrence $\mathcal{C}$. 

The rotational invariance of the reduced two-spin state, given by $\hat{\rho} = p |S\rangle\langle S| + (1-p)\,\hat{\mathcal{I}}/4$, allows to express both $\mathcal{C}$ and $\mathcal{C}_{\rm GME}$ as a function of $p$, the excess probability associated with the singlet state $|S\rangle$. 
As a result, we find:
\begin{gather}\label{monogamy}
    \mathcal{C}_{\rm GME} = \left\{\frac{3}{2} \left[ 1-\frac{1}{9} (1+2\,\mathcal{C}_{\rm max})^2 \right]\right\}^{1/2}\;,
\end{gather}
where $\mathcal{C}_{\rm max}$ denotes the maximum of the two-spin concurrence among all the possible spin pairs.
Equation (\ref{monogamy}) formalizes the complementarity between spin-pair and GME in the spin states obtained through the QFT, and in all the cases considered hereafter, where the subsystem with the lowest linear entropy ($A_m$) corresponds to a spin pair. Besides, from the monogamy of entanglement \cite{Coffman00a} we derive an upper bound for $\mathcal{C}_{\rm max}$ in cyclic systems: this, combined with Eq. (\ref{monogamy}), results in a lower bound for the GME concurrence: $\mathcal{C}_{\rm GME}\ge (1-\sqrt{2}/3)^{1/2}$ (see Sec. V of the Supplemental Materials). This further supports the generality of the derived results on the abundant presence of GME in this class of systems.

In order to further characterize the considered states, we note that the $n$BRDMs correspond, for $n=N$, to pure spin states, and specifically to spin singlets. These states can be expressed as linear superpositions of resonating valence bond states, {\it i.e.} products of $N/2$ two-spin singlets delocalized on different pairs of orbitals $\phi_i$. More specifically, all their entanglement properties (concurrences and linear entropies) coincide with those of the ground states of antiferromagnetic Heisenberg rings, which have been investigated by different means \cite{Guhne05a,Guhne09a,Troiani11a,Siloi12a,Siloi14a}.

The extraction of GME through the QFT can be further generalized. In fact, in Eq. (\ref{Psi in terms of d}) the relation between the $N/2$ doubly-occupied orbitals $\varphi_1,\dots,\varphi_{N/2}$ and the $N$ extraction orbitals $\phi_i$ is defined by the first $N/2$ lines of the QFT ($u_{jk}$ with $j=1,\dots,N/2$). A high or maximal degree of GME can be generally obtained by using other sets of $N/2$ lines, identified by the indices $j=l_1,\dots,l_{N/2}$. In fact, the numerical calculations performed for $4 \le n \le 10$ systematically show that $\mathcal{C}_{\rm GME}$: takes the same value for any set of $N/2$ consecutive lines, including the case where these coincide with the first ones ($l_i=i$); vanishes if the $N/2$ lines are the even- ($l_i=2i$) or odd-numbered ($l_i=2i-1$) ones; is maximal ($\mathcal{C}_{\rm GME}=1$) in most of the other cases (see Sec. V of the Supplemental Material \cite{SM}). Altogether, these results show the possibility of extracting a maximal degree of GME from simple non-correlated states.
}

\begin{figure}
\centering
\includegraphics[width=0.42\textwidth]{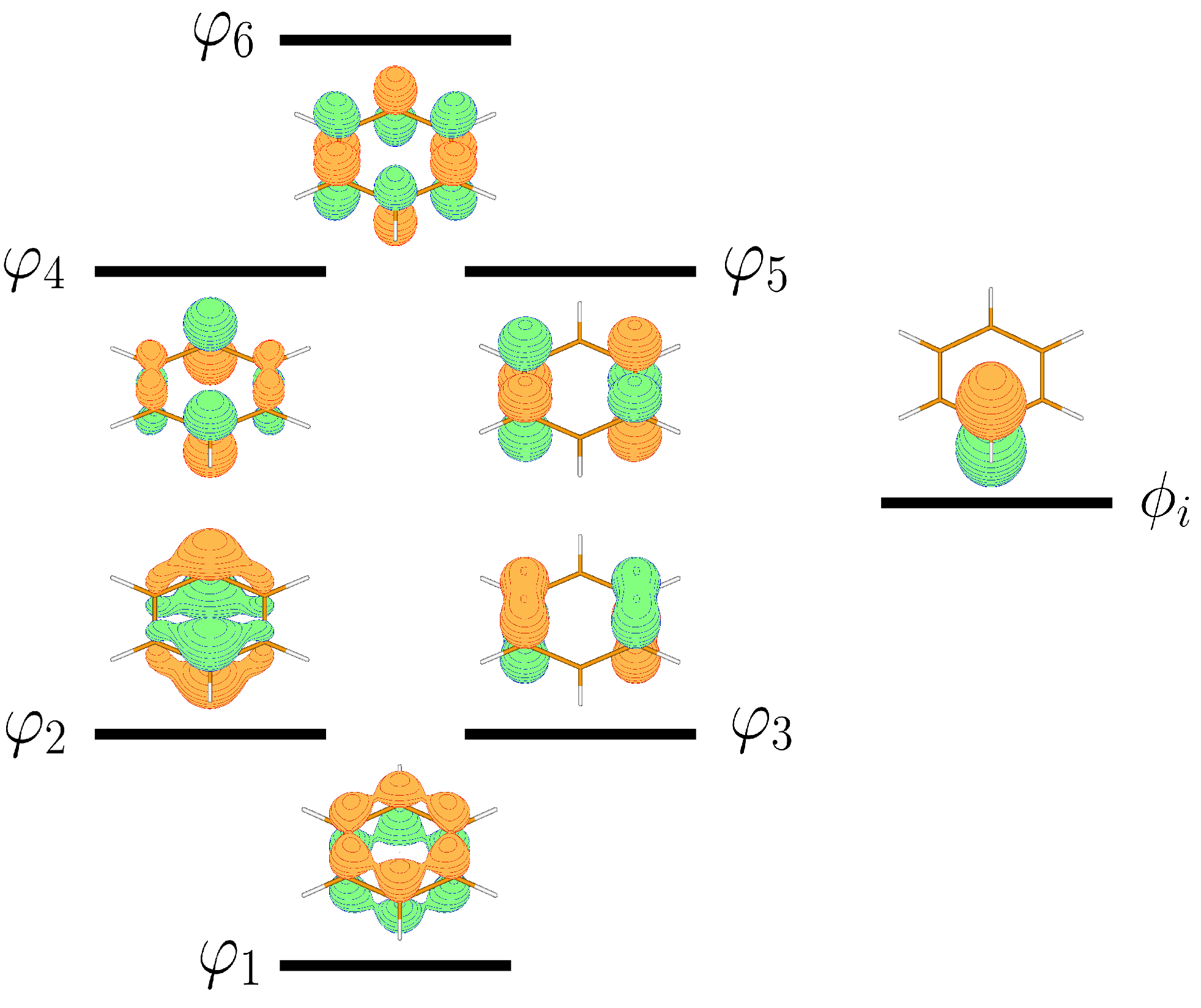}
\caption{Delocalized and localized orbitals in benzene: on the left part, the six $\{\varphi_i\}$ delocalized MOs (natural orbitals of the CASSCF wave function); on the right, one of the six (equivalent by symmetry) $\{\phi_i\}$ localized OAO. The horizontal black lines indicate the orbital energies. The localized orbitals all have the same energy expectation value.
}
\label{fig3}
\end{figure}

{\color{black} {\em GME from the Hubbard model}.~ In order to extend the scope of the above results, let us also consider  non-trivial correlated eigenstates of the Hubbard model \cite{Hubbard2013}. In a ring-shaped system at half-filling, the number of electrons ($N$) coincides with the number of sites. The extraction of $n=N$ particles takes place from the state $|\Psi\rangle$, which is identified with the ground state of the Hubbard Hamiltonian, and each orbital $\phi_i$ is localized at the $i$-th site. Quite remarkably, for all the considered values of $n$, the $n$BRDM obtained from the ground state of the Hubbard model coincides with that obtained from the noninteracting state [Eq. (\ref{eqn:Psi})] through the QFT ($U=U_{QFT}$, with $l_i=i$). 
This coincidence has been numerically verified for $n = 4,6,8,10$ (Table \ref{tab:my_label}), and can be reasonably expected to hold also for $n > 10$. 

Furthermore, the $n$BRDMs derived from the ground states of the Hubbard model for different values of $n$ coincide with the ground states of the antiferromagnetic Heisenberg ring. Interestingly, this happens not only in the limit $|t|\ll U$ (where such coincidence is expected), but in all the considered range of parameters ($5 \le U/|t| \le 50$). 

The presence of GME is not an exclusive feature of the ground states. For example, let us consider the first nondegenerate (singlet) excited state of the Hubbard model. The resulting $n$BRDM displays a maximal GME concurrence ($\mathcal{C}_{\rm GME}=1$) and no spin-pair entanglement. As for the ground state, the same entanglement properties are obtained for the first excited singlet state of the antiferromagnetic Heisenberg ring (see Sec. V of the Supplemental Material \cite{SM}).
}

{\color{black} \em GME from atomistic systems}.~ 
The transformations $U$ considered so far have related two unspecified {\color{black} (or structureless, in the case of the Hubbard model)} sets of orbitals. {\color{black} Turning to realistic} quantum states of matter, {\color{black} it is desirable that both} the orbitals $\varphi_i$ and $\phi_j$ possess a clear physical meaning, {\color{black} possibly derived from a realistic, {\em ab-initio} many-body Schr\"odinger equation.}
For example, the $\varphi_i$ may be the set of orbitals that allow the approximation of the system ground state with a single Slater determinant (as in the Hartree-Fock approximation) or, when necessary, the inclusion of nontrivial Coulomb-interaction effects, via an appropriate correlated {\em ansatz}. The $\phi_j$, instead, should correspond to a set of localized orbitals, which justify in principle the interpretation of the $n$BRDM as the spin state of an extracted $n$-particle system.

{\color{black} To provide concrete examples, we investigate the singlet ground states of a number of molecules, characterized by different sizes and geometries. In all cases, $n$ coincides with the number of electrons in the active space of the Complete Active Space Self Consistent (CASSCF) wave function, and $ N - n > 0 $ is the number of core electrons. Only the former ones affect the expression of the $n$BRDM, because both the orbitals $\varphi_i$ and $\phi_j$ belong to the active space.}

We start by considering the case of benzene {\color{black} (C$_6$H$_6$), which includes $n=6$ electrons belonging to the active space and 36 electrons frozen in doubly-occupied core orbitals ($N=42$).} The CASSCF calculations provide six delocalized $\pi$ Molecular Orbitals (MOs) $\varphi_i$, and the $n \times n$ unitary transformation $U$ is defined so as to rotate them into the six orthogonal atomic orbitals (OAOs)~\cite{Genesis,JCE-OVB} $\phi_j$ that maximize the overlap with the $2p_z$ orbitals of the C atoms (Fig. \ref{fig3}). {\color{black} This transformation $U$ shares with the QFT a qualitative feature that plays a fundamental role in the generation of GME, namely the fact that each of the orbitals $\varphi_i$ is expressed as a combination, with comparable weights, of all of the $\phi_j$ (see Sec. I of the Supplemental Material \cite{SM})}. 

\begin{table}[]
    \centering
    \begin{tabular}{|c|c|c|c|c|c|}
\hline
{\color{black} Hubbard ring} & {\color{black} $N$} & {\color{black} $n$} & {\color{black} $\mathcal{C}_{\rm GME} (|\Psi_s^{(n)}\rangle)$} & {\color{black} $\mathcal{C}(\rho_{A_m})$} & {\color{black} $A_m$} \\
\hline 
        {\color{black} 4  sites (GS)} &  {\color{black} 4} &  {\color{black} 4} & {\color{black} 0.913} & {\color{black} 0.500} & {\color{black} $\{s_i,s_{i+1}\}$} \\
        {\color{black} 6  sites (GS)} &  {\color{black} 6} &  {\color{black} 6} & {\color{black} 0.958} & {\color{black} 0.434} & {\color{black} $\{s_i,s_{i+1}\}$} \\
        {\color{black} 6  sites (ES)} &  {\color{black} 6} &  {\color{black} 6} & {\color{black} 1.000} & {\color{black} 0.000} & {\color{black} $\{s_i,s_{j}\}$} \\
        {\color{black} 8  sites (GS)} &  {\color{black} 8} &  {\color{black} 8} & {\color{black} 0.973} & {\color{black} 0.411} & {\color{black} $\{s_i,s_{i+1}\}$} \\
        {\color{black} 10 sites (GS)} & {\color{black} 10} & {\color{black} 10} & {\color{black} 0.978} & {\color{black} 0.403} & {\color{black} $\{s_i,s_{i+1}\}$} \\
\hline
Molecule & $N$ & $n$ & $\mathcal{C}_{\rm GME} (|\Psi_s^{(n)}\rangle)$ & $\mathcal{C}(\rho_{A_m})$ & $A_m$ \\
\hline
        Benzene (GS)      & 42 & 6 & 0.958 & 0.434 & $\{s_i,s_{i+1}\}$ \\
        {\color{black} Benzene (ES)} & {\color{black} 42} & {\color{black} 6} & {\color{black} 1.000} & {\color{black} 0.000} & {\color{black} $\{s_i,s_{j}\}$} \\
        Hexatriene (GS) & 44 & 6 & 0.300 & 0.955 & $\{s_{1},s_{2}\}$ \\
        Decapentaene (GS) & 72 & 10 & 0.357 & 0.936 & $\{s_{1},s_{2}\}$ \\
\hline
    \end{tabular}
    \caption{{\color{black} Genuine multipartite entanglement ($\mathcal{C}_{\rm GME}$) and two-spin  concurrences ($\mathcal{C}$) obtained for the Hubbard rings and for different molecules (GS and ES denote the ground and excited states, respectively). The results given for the Hubbard model coincide with those obtained through the QFT, with the same value of $n=N$ (and with $l_i=i$). The subsystem $A_m$ is the one whose reduced density operator $\rho_{A_m}$ displays the lowest linear entropy, and thus determines the value of the GME concurrence. In the case of the molecules, $n$ and $N$ are the number of electrons in the active space and the overall number of electrons, respectively.}}
    \label{tab:my_label}
\end{table}
 
{\color{black} All the entanglement properties of the benzene ground state, including the nearly maximal value of the GME concurrence, coincide with those obtained from the single Slater determinant through the $U_{QFT}$ and from the ground state of the Hubbard ring in the case $n=6$ (Table \ref{tab:my_label}). This shows that the normalized projection of the benzene ground state onto the subspace characterized by a single occupation of all the $\phi_i$ orbitals, coincides with that of the Hubbard ring. The correspondence between the benzene molecule and the Hubbard ring is not limited to the ground state, but extends to the first excited singlet, from which we obtain the same values of the spin-pair and GME concurrences.}

In order to gain further insight into the relation between spin entanglement {\color{black} and symmetry, we compare the above case of a cyclic system (benzene) with that of two linear molecules: hexatriene (C$_{6}$H$_{8}$) and decapentaene (C$_{10}$H$_{12}$).} In both cases, starting from the multi-configurational CASSCF ground state $|\psi\rangle$ and identifying the $\phi_i$ with the OAOs, we obtain a genuinely multipartite entangled spin state. However, the values of $\mathcal{C}_{\rm GME}$ are significantly lower than that obtained for benzene (Table \ref{tab:my_label}). {\color{black} This results from a partial dimerization of the chain, which reflects the alternation of single and double bonds along the chain, and is particularly pronounced at the edges of the molecule. For all the considered molecules, the minimum linear entropy, which determines the value of $\mathcal{C}_{\rm GME}$, is obtained for the subsystems consisting of the pairs of neighboring spins: the relation between the GME and the highest spin-pair concurrence is thus given by Eq. (\ref{monogamy}).}

{\color{black} {\em Conclusions}.~
 We have shown that abundant genuine multipartite entanglement (GME) can be extracted from a variety of many-electron systems and states: from ground to excited states and from non-interacting to interacting systems.

We have demonstrated that, contrary to common wisdom, maximal GME can be extracted also from the case of single-reference states, specifically through orbital transformations defined by the quantum Fourier transform. 

By considering prominent models of electronic structure of matter --- Hubbard  and {\em ab initio} atomistic models --- we have shown that close to maximal GME can also be extracted from correlated ground and excited states, through physically motivated localized orbitals. We have found a remarkable, quantitative agreement between the entanglement properties derived from different models that share a common cyclic symmetry.
 
Through {\em ab initio} calculations, we have finally revealed system-dependent features that can affect the extraction of GME. In particular, we have found that linear molecules (hexatriene and decapentaene) present a pronounced dimerization of the ground state, reflected, at the chemical level, in the alternation of single and double bonds, which strongly limits the amount of GME.

Because superselection rules are consistently fulfilled in our analyses, the extracted entanglement can in principle be available as a resource. Thus, potential developments at the experimental and technological levels are anticipated.}

\acknowledgments
The authors acknowledge financial support from the Ministero dell’Universit\`a e della Ricerca (MUR) under the Project PRIN 2022 number 2022W9W423 and the PNRR Project PE0000023-NQSTI.  


%

\end{document}